\documentclass[pra,twocolumn,superscriptaddress,showpacs,amsmath,amssymb]{revtex4}

\usepackage{graphicx}
\usepackage{dcolumn}
\usepackage{bm}
\usepackage{epsfig}
\usepackage{amsmath}
\usepackage{amssymb}
\usepackage{color}
\usepackage{bbm}
\usepackage{CJK}
\usepackage[colorlinks=true,citecolor=blue,linkcolor=blue,urlcolor=blue]{hyperref}
\begin{document}
\begin{CJK*}{UTF8}{gbsn}
\title{Work statistics in non-Hermitian evolutions with Hermitian endpoints}
\author{Zheng-Yang Zhou}
\altaffiliation[zhengyang.zhou@riken.jp]{}
\affiliation{Theoretical Quantum Physics Laboratory, RIKEN Cluster for Pioneering Research, Wako-shi, Saitama 351-0198, Japan}
\affiliation{Quantum Physics and Quantum Information Division, Beijing Computational Science Research Center, Beijing
100094, China}
\author{Ze-Liang Xiang}
\altaffiliation[xiangzliang@mail.sysu.edu.cn]{}
\affiliation{School of Physics, Sun Yat-sen University, Guangzhou 510275, China}
\author{J. Q. You}
\altaffiliation[jqyou@zju.edu.cn]{}
\affiliation{Interdisciplinary Center of Quantum Information, State Key Laboratory of Modern Optical Instrumentation, and Zhejiang Province Key Laboratory of Quantum Technology and Device, Department of Physics, Zhejiang University, Hangzhou 310027, China}
\affiliation{Beijing Computational Science Research Center, Beijing
100094, China}
\author{Franco Nori (野理)}
\altaffiliation[fnori@riken.jp]{}
\affiliation{RIKEN Center for Quantum Computing (RQC), Wako-shi, Saitama 351-0198, Japan}
\affiliation{Theoretical Quantum Physics Laboratory, RIKEN Cluster for Pioneering Research, Wako-shi, Saitama 351-0198, Japan}
\affiliation{Physics Department, the University of Michigan, Ann Arbor, Michigan 48109-1040, USA}
\date{\today}

\begin{abstract}
Non-Hermitian systems with specific forms of Hamiltonians can exhibit novel phenomena. However, it is difficult to study their quantum thermodynamical properties. In particular, the calculation of work statistics can be challenging in non-Hermitian systems due to the change of state norm. To tackle this problem, we modify the two-point measurement method in Hermitian systems. The modified method can be applied to non-Hermitian systems which are Hermitian before and after the evolution. In Hermitian systems, our method is equivalent to the two-point measurement method. When the system is non-Hermitian, our results represent a projection of the statistics in a larger Hermitian system. As an example, we calculate the work statistics in a non-Hermitian Su-Schrieffer-Heeger model. Our results reveal several differences between the work statistics in non-Hermitian systems and the one in Hermitian systems.
\end{abstract}
\maketitle
\end{CJK*}


%
%


\section{Introduction}

Ideal physical systems are conceptually Hermitian, but realistic systems are sometimes non-Hermitian because of their interactions with their environments. Non-Hermitian systems occur in various fields of physics and are experimentally accessible~\cite{PTbook1,genhsys1,genhsys2,nhsys1,nhsys2,nhsys3,nhsys4,nhsys5,nhsys6,nhsys7}. Many fascinating phenomena related to non-Hermiticity were discovered in, e.g., topological systems~\cite{nhtopo1,nhtopo2,nhtopo3,nhtopo4,nhtopo5}, many-body systems~\cite{nhmb1,nhmb2}, adiabatic passage~\cite{nhap1,nhap2,nhap3,nhap4,nhap5,nhap6}, nonreciprocal scattering~\cite{nhnonre1,nhnonre2,nhnonre3}, and localization-delocalization transitions~\cite{nhlocal1,nhlocal2,nhlocal3,nhlocal4}. Many works have introduced non-Hermiticity to well-known systems, especially those already shown to have novel properties in the Hermitian cases. Among these systems, the non-Hermitian Su-Schrieffer-Heeger (SSH) model~\cite{SSHmodel1,SSHmodel2,SSHmodel3,SSHmodel4,expnhSSH} plays an important role, since it exhibits both a PT (parity-time)-symmetry-breaking phase transition and a topological phase transition. However, these non-Hermitian systems also bring problems which are not considered in Hermitian quantum mechanics.

One important problem is the quantum thermodynamical description of non-Hermitian systems. Previous works in quantum thermodynamics~\cite{quanthermre1,quanthermre2,quanthermre3,quanthermre4,quanthermre5} mainly considered Hermitian systems, so that the norm of the state is usually assumed to be conserved. However, such assumption is not always valid in non-Hermitian systems, which can cause troubles in specific thermodynamics studies, e.g., work statistics~\cite{jequal1,tpm1,tpm2,tpm3,npmw1,npmw2,npmw3,wwco1,wwco2,wopen}. Along with the typical complex-eigenvalue problem in non-Hermitian systems, two main problems can arise in non-Hermitian work statistics. First, many quantum thermodynamics results are based on different kinds of trajectories, which refer to processes with certain probability. The definition of trajectories in non-Hermitian systems becomes ambiguous because the probability of a trajectory usually changes with time. Second, the change of the state norm can give rise to entropy changes, so that we need to distinguish work from heat flow. These problems can be solved by introducing a bi-orthogonal basis if all the eigenvalues are real~\cite{biorthermo1,biorthermo2}, but the problems become complicated when the system passes through an exceptional point (EP). Thus, a more general method is necessary to understand the thermodynamic work in non-Hermitian systems.

Here, we study the work statistics in non-Hermitian systems which are Hermitian at the initial and the final times. To have a good definition of energy change, we consider systems with non-Hermitian Hamiltonians only during their evolution. Our main goal is to solve the trajectory problem and the entropy problem caused by state norm change. To avoid these two problems, we purify the state of the system and define the statistics in an enlarged Hilbert space. After the state purification, the work statistics can be defined in a non-Hermitian system. Our method is then compared to the ordinary two-point measurement method, and is shown to be a projection of Hermitian work statistics. As an example to illustrate the properties of non-Hermitian work statistics, we calculate the work statistics in the non-Hermitian SSH model. The distributions of work on the two sides of the EP are compared. We also introduce some additional non-Hermitian terms to show that work statistics can be significantly modified by a negligible variation in the spectrum.

This paper is organized as follows. In Sec.~\ref{method}, we summarize the difficulties in non-Hermitian work statistics, and present our method. In Sec.~\ref{workstatiscs}, as an example ,the work statistics of the non-Hermitian SSH model is derived. Section~\ref{conclusion} presents our conclusions.
\section{Work statistics in non-Hermitian systems}
\label{method}
\subsection{Model considered}
\begin{figure}[t]
\center
\includegraphics[width=3in]{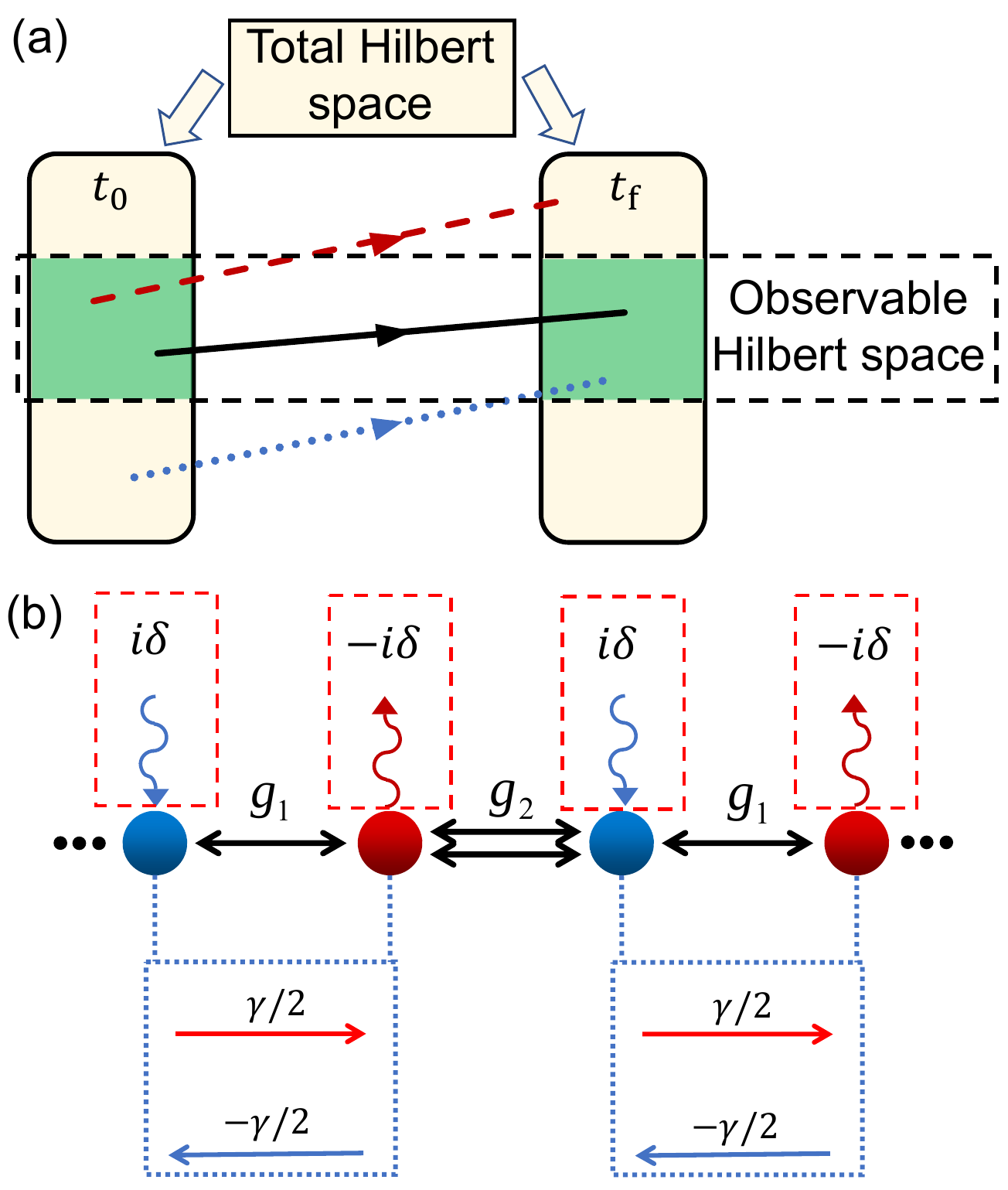}
\caption{(a)~Schematic illustration of the statistics in non-Hermitian systems. The processes with decreasing norm, conserved norm, and increasing norm are shown with a red dashed line, a black solid line, and a blue dotted line, respectively. (b)~Illustration of the Hermitian Su-Schrieffer-Heeger model and two types of non-Hermitian terms. The terms in the red dashed squares and the terms in the blue dotted squares represent the loss-gain terms and non-reciprocal terms, respectively.}
\label{fig1}
\end{figure}

We consider a non-Hermitian system which consists of a Hermitian part $H_{0}=H_{0}^{\dag}$ and a non-Hermitian part $H_{\rm nh}\neq H_{\rm nh}^{\dag}$. The non-Hermitian part is assumed to vanish at the beginning time $t=0$ and the ending time $t=t_{\rm f}$. With this assumption, the energy and the thermal equilibrium state can be well defined at these two time points. The system is assumed to be in the thermal equilibrium state $\rho_0=\exp(-\beta H_0)/{\rm Tr}\{\exp(-\beta H_0)\}$ at the beginning time, where $\beta$ is the inverse of the temperature. This thermal state is achieved by coupling to an environment, but we assumed the coupling strength to be negligible during the following evolution. When the non-Hermitian part is on, the system evolves according to the following relation (setting $\hbar=1$)~\cite{nhevo}:
\begin{equation}
\rho(t)=\frac{U(t)\rho_0U^{\dag}(t)}{N_{\rho}(t)},
\label{nhevorho}
\end{equation}
where $U(t)\equiv T_+\exp\left(-i\int_0^tds[H_{\rm nh}(s)+H_0]\right)$ is the evolution operator, $N_{\rho}(t)\equiv{\rm Tr}\left\{U(t)\rho_0U^{\dag}(t)\right\}$ is the normalization factor, and the term $T_+$ is time ordered matrix. Note that the approach in Eq.~(\ref{nhevorho}) describes the non-Hermiticity of open systems, which is different from the one for closed non-Hermitian systems~\cite{biorthermo1,biorthermo2,nhevo2} (also see Appendix A). To focus on the work related to non-Hermitian process, we assume $H_{0}$ to be time independent. This dynamics corresponds to the projection of the system evolution onto a set of incomplete basis, as shown in Fig.~\ref{fig1}(a). Although the total system is in a Hermitian space, we can observe it in a smaller state space and describe the dynamics with the non-Hermitian evolution in Eq.~(\ref{nhevorho}). In spite of the dynamical equivalence, this description brings troubles to the work statistics in non-Hermitian systems. Both theoretical~\cite{tpm1,tpm2,tpm3,npmw1,npmw2,npmw3,npmw3,wwco1} and experimental~\cite{wfexp1,wfexp3} approaches for studying work statistics depend on trajectories, which describe physical or virtual processes with certain properties. The probabilities of trajectories are not conserved in non-Hermitian systems, which makes the definition of trajectories difficult in these systems. Now, we discuss related problems and the way to solve these.
\subsection{Difficulties of applying the two-point measurement method}
The two-point measurement method~\cite{tpm3,npmw1,quanthermre3,wopen} is a widely used tool in work statistics. However, such a method cannot be directly applied to non-Hermitian systems. Therefore, we analyze the problems of the two-point measurement method in non-Hermitian cases, so that the necessary modifications can be introduced.

For the two-point measurement in Hermitian problems, the system is initially prepared in the thermal equilibrium state $\rho_0=\exp(-\beta H(0))/{\rm Tr}\{\exp(-\beta H(0))\}$ with the Hamiltonian $H(0)$ and the inverse of temperature $\beta$. The first projection measurement is implemented on the thermal state according to the eigenstate basis of the system. After the measurement, we have the probability $P_n=\exp(-\beta E_{n}(0))/{\rm Tr}\{\exp(-\beta H(0))\}$ to obtain the eigenstate $|n(0)\rangle$ with eigenvalue $E_{n}(0)$. The collapsed state $|n(0)\rangle$ then evolves under the influence of the time-dependent Hamiltonian
\begin{equation}
|\psi_n(t)\rangle=T_+\exp\left(-i\int_0^t\!\!ds\;H(s)\right)|n(0)\rangle\equiv U(t)|n(0)\rangle.
\end{equation}
After the evolution, the second projection measurement is implemented on the system according to some measurement basis $|m(t)\rangle$ (usually the eigenstate basis of the system at the beginning time or the ending time), which gives the state $|m(t)\rangle$ with the probability
\begin{eqnarray}
T_{n,m}&=&|\langle m(t)|\psi_n(t)\rangle|^2.
\end{eqnarray}
The probability to observe the whole process mentioned above is $P_{n,m}=T_{n,m}P_{n}$. Note that the work done in this process is $W_{n,m}=E_{m}(t)-E_{n}(0)$, where $E_{m}(t)$ and $E_{n}(t)$ are the energies of $|m(t)\rangle$ and $|n(0)\rangle$, respectively. Finally, the quantity $W_{n,m}$, which corresponds to a stochastic process with the distribution $P_{n,m}$, is used to describe the work statistics. This definition of work has two main advantages. First, the average value of work is equivalent to the classical work. Second, the Jarzynski relation~\cite{jequal1}, which is initially a classical extension of the dissipation-fluctuation relation, can be satisfied in closed quantum systems and open quantum systems at thermal equilibrium under this two-point measurement scheme~\cite{quanthermre2,quanthermre3,wopen}.

Now we show that the two-point measurement method cannot be directly applied to non-Hermitian systems. The first problem is that the system dynamics can be changed by the projection measurement.
In Hermitian systems, if we take the average of the states $|\psi_n(t)\rangle$ according to the probability $P_n$, the density matrix without the first measurement can be retained,
\begin{eqnarray}
U(t)\rho_0U^{\dag}(t)=\sum_nP_n|\psi_n(t)\rangle\langle\psi_n(t)|.
\end{eqnarray}
As a result, the first measurement does not change the dynamics of the system on average. Thus, this approach can reveal the thermodynamic properties of the process. However, the situation becomes complicated for non-Hermitian systems. A non-Hermitian system evolves according to Eq.~(1) with
a normalization factor $N_{\rho}(t)$ which depends on the initial state $\rho_0$. Therefore, different $|n(0)\rangle$'s experience different dynamics. For example, assume that the first measurement gives the result $|n(0)\rangle$. The corresponding state at $t$ becomes
\begin{eqnarray}
\rho(t)=\frac{U(t)|n(0)\rangle\langle n(0)|U^{\dag}(t)}{\langle n(0)|U^{\dag}(t)U(t)|n(0)\rangle}.
\end{eqnarray}
In a non-Hermitian system, the product $U^{\dag}(t)U(t)$ is usually not an identity operator, which contributes a normalization factor depending on the result of the first measurement. If we average over the results of the first measurement, the obtained state is
\begin{eqnarray}
\tilde{\rho}(t)=\sum_nP_n\frac{U(t)|n(0)\rangle\langle n(0)|U^{\dag}(t)}{\langle n(0)|U^{\dag}(t)U(t)|n(0)\rangle},
\end{eqnarray}
instead of the density matrix without the measurement. As a result, the two-point measurement approach provides the work statistics of another process. Such effects are quite common in systems with initial coherence (see e.g.~\cite{wwco1,wwco2}), but the problem here originates from the non-preservation of the norm in non-Hermitian systems~\cite{nhevo2}.

The second problem is the purity change due to the change of the norm. In general, non-Hermitian dynamics can influence the system purity (also the entropy)~\cite{nhevo,nhwec}. However, the work has no impact on the entropy of a closed system because the corresponding process is reversible. Therefore, it is straightforward to expect that the non-Hermiticity can affect both work and heat current.

\subsection{Work statistics with purification}
As mentioned above, the work statistics based on the two-point measurement has problems in non-Hermitian systems. To solve these problems, we introduce some modifications to the two-point measurement method, and present our way of calculating work statistics in non-Hermitian systems.

Although the purity of the system usually changes under the influence of a non-Hermitian Hamiltonian, non-Hermitian dynamics preserves the purity of a pure initial state. For such initial states, there is no entropy change, because entropy is always zero for a pure state. In such a case, we can safely calculate the work with an energy difference. A thermal state with the inverse temperature $\beta$ can be expressed as a pure state in a larger Hilbert space if the eigenstate thermalization hypothesis is assumed~\cite{esther1,esther2}. In the following part, we call {\it the Hilbert space in addition to our system} ``heat bath'' (also see Appendix B). However, this heat bath is different for a large system, e.g., harmonic oscillators, in thermal equilibrium state. By assuming our system initial state to be the result of eigenstate thermalization, it can be expressed as the partial trace of a pure state:
\begin{eqnarray}
\rho_0={\rm Tr}_{\rm bath}\{|\Psi_{\rm tot}\rangle\langle\Psi_{\rm tot}|\},\nonumber
\end{eqnarray}
with
\begin{eqnarray}
|\Psi_{\rm tot}\rangle&=&\sum_nC_n|n(0)\rangle\otimes|\psi_{n}^{\rm bath}\rangle,\nonumber\\
\delta_{n,m}&=&\langle\psi^{\rm bath}_{n}|\psi_{m}^{\rm bath}\rangle.\label{purification}
\end{eqnarray}
Note that $|\psi^{\rm bath}_{n}\rangle$ is not necessarily an eigenstate of the bath. We further assume the following relations:
\begin{eqnarray}
&&C_n=\sqrt{\frac{\exp(-\beta E_{n}(0))}{{\rm Tr}\{\exp(-\beta H_0)\}}},\nonumber\\
&&E_{n}^{\rm bath}-E_{m}^{\rm bath}=E_{m}(0)-E_{n}(0),\nonumber\label{bathbasis}
\end{eqnarray}
and
\begin{eqnarray}
&&\langle\psi_{n}^{\rm bath}|H_{\rm bath}|\psi^{\rm bath}_{m}\rangle=0,~~{\rm for}~~m\neq n.\label{assumptions}
\end{eqnarray}
Here, $E_{n}(0)$ is the eigenvalue of $|n(0)\rangle$, $\beta=1/(k_{\rm B}T)$ is the inverse of the bath temperature, $H_{\rm bath}$ is the Hamiltonian of the bath, and $E^{\rm bath}_{n}\equiv\langle\psi_{n}^{\rm bath}|H_{\rm bath}|\psi_{n}^{\rm bath}\rangle$ is the average energy of the state $|\psi_{n}^{\rm bath}\rangle$.

Next, we briefly justify these assumptions. The first equation in Eq.~(\ref{assumptions}) makes sure that the reduced state in the system's Hilbert space is the thermal state with temperature $T$. The second equation in Eq.~(\ref{assumptions}) refers to the eigenstate thermalization assumption, which is a possible thermalization mechanism in closed systems. The bath energies $E_{n}^{\rm bath}$ and $E_{m}^{\rm bath}$ in the second equation of Eq.~(\ref{assumptions}) are in general not eigenvalues but average values. However, note that $|\psi_{n}^{\rm bath}\rangle$ is the superposition of eigenstates with eigenvalues close to $E_{n}^{\rm bath}$, which is important to the last assumption. Let us assume that $|\psi_{n}^{\rm bath}\rangle$ contains eigenstates with eigenvalues within the range
$$(E_{n}^{\rm bath}-\delta E_{n}^{\rm bath},E^{\rm bath}_{n}+\delta E^{\rm bath}_{n}).$$
The last assumption is fulfilled if we have
$$|E_{n}^{\rm bath}-E_{m}^{\rm bath}|>|\delta E_{n}^{\rm bath}|+|\delta E^{\rm bath}_{m}|.$$
The last assumption is necessary to avoid the system-energy shift caused by the bath. {For example, terms like $[(|n(0)\rangle+|m(0)\rangle)\langle k(0)|+h.c.]$ can drive both $|n(0)\rangle$ and $|m(0)\rangle$ to a third state $|k(0)\rangle$, which provides a state in the ``system + bath'' Hilbert space with the following form,}
\begin{eqnarray}
|k(0)\rangle\otimes(|\psi_{n}^{\rm bath}\rangle+|\psi_{m}^{\rm bath}\rangle).
\end{eqnarray}
If the coherence of the bath provides energy, the energy {change of this transition} can be modified by this energy. This effect is similar to the case of exchange energy. Apparently, an ordinary thermal bath does not have such an effect, because the system is assumed to be fully described by the reduced density matrix. Therefore, the coherence of the bath states does not contribute to the bath energy.

Based on these assumptions, the work statistics of the non-Hermitian system can be studied. Now, we turn on the non-Hermitian Hamiltonian, so that the system evolves according to Eq.~(1). This evolution can also be written in the Hilbert space of the system plus the bath,
\begin{eqnarray}
\rho_{\rm tot}(t)&=&\frac{U(t)U_{\rm bath}(t)|\Psi_{\rm tot}\rangle\langle\Psi_{\rm tot}|U^{\dag}(t)U^{\dag}_{\rm bath}(t)}{N_{\rho_{\rm tot}}(t)},\nonumber\label{nhtotdynamics}
\end{eqnarray}
with
\begin{eqnarray}
U_{\rm bath}(t)=\exp\left(-iH_{\rm bath}t\right).
\end{eqnarray}
Unlike the Hermitian Hamiltonian, the non-Hermitian Hamiltonian can also have an effect on the heat bath. To show this effect, we trace out the system degrees of freedom in Eq.~(\ref{nhtotdynamics}),
\begin{eqnarray}
\rho_{\rm bath}(t)&=&{\rm Tr}_{\rm sys}\{\rho_{\rm tot}(t)\}\nonumber\\
                  &=&\sum_n P_{{\rm bath},n}(t) |\psi_{n}^{\rm bath}(t)\rangle\langle\psi_{n}^{\rm bath}(t)|,\nonumber
\end{eqnarray}
with
\begin{eqnarray}
P_{n}^{\rm bath}(t)&=&\frac{C_n^2\langle n(0)|U^{\dag}(t)U(t)|n(0)\rangle}{N_{\rho_{\rm tot}}(t)},\nonumber
\end{eqnarray}
and
\begin{eqnarray}
|\psi_{n}^{\rm bath}(t)\rangle=U_{\rm bath}(t)|\psi_{n}^{\rm bath}\rangle.
\end{eqnarray}
For a non-Hermitian system, the term $\langle n(0)|U^{\dag}(t)U(t)|n(0)\rangle$ usually changes with time, so the bath dynamics is also influenced by the non-Hermitian Hamiltonian of the system. Therefore, generating a non-Hermitian Hamiltonian contains operations outside the Hilbert space of the system.

As a result, the work done on the bath should also be considered. However, it is impossible to implement work statistics in the Hilbert space of the system plus the bath, because the bath states here are, in general, not the eigenstates of the bath Hamiltonian. Instead of the eigenstates of the bath, we estimate the work on the bath with the basis formed by $|\psi^{\rm bath}_{n}(t)\rangle$. Note that these states are complete and orthogonal for the problem studied here. In addition, by considering the bath energy, all the possible results of the first measurement have the same energy, which indicates that the first measurement is not necessary. Therefore, we do not have problems related to the projection measurement.
We start from the initial state $|\Psi_{\rm tot}\rangle$ with the energy $E_{\rm tot}=E_{n}(0)+E^{\rm bath}_{n}$. After the influence of the non-Hermitian Hamiltonian, we measure the total system on the basis,
\begin{eqnarray}\label{nhworkstatistics:finalbasis}
|m({0})\rangle\otimes|\psi^{\rm bath}_{n}(t_{\rm f})\rangle.
\end{eqnarray}
It is straightforward to see that the energy difference between $|m(0)\rangle\otimes|\psi^{\rm bath}_{n}(t_{\rm f})\rangle$ and $|\Psi_{\rm tot}\rangle$ is just
\begin{eqnarray}
E_{m}(0)+E^{\rm bath}_{n}-E_{\rm tot}=E_{m}(0)-E_{n}(0).
\end{eqnarray}
The probability of the transition from $|\Psi_{\rm tot}\rangle$ to $|m(0)\rangle\otimes|\psi_{n}^{\rm bath}(t_{\rm f})\rangle$ is
\begin{eqnarray}
P_{m,n}&=&\langle m({0})|\otimes\langle\psi_{n}^{\rm bath}(t_{\rm f})|\rho_{\rm tot}(t_{\rm f})|m({0})\rangle\otimes|\psi_{n}^{\rm bath}(t_{\rm f})\rangle.\nonumber\\
\end{eqnarray}
Since the system has a unit probability to be in the state $|\Psi_{\rm tot}\rangle$, the characteristic function of the work can be calculated with
\begin{eqnarray}
\chi(u)=\sum_{m,n}\exp({iu[E_{m}(0)-E_{n}(0)]})P_{m,n}.\label{workchf}
\end{eqnarray}
Equation~(\ref{workchf}) is similar to the ordinary characteristic function of work in Hermitian systems, but it has different meanings. The indexes $n$ and $m$ correspond to two Hilbert spaces in one measurement instead of two successive measurements; what is calculated in our method is not the work done on the system but the work done on the system plus the bath.

System work statistics containing measurements on the heat bath might be counter-intuitive. However, such definition is necessary even in Hermitian cases if the system-bath coupling exists~\cite{wopen}. Although there is no direct coupling in our case, measurements on the system can also influence the bath.
\subsection{Relation with the work statistics in Hermitian systems}
We now discuss the relation between the work statistics in our work and the ordinary two-point measurement method in Hermitian systems. Here we mainly focus on two issues. One is Hermitian limit of our method, the other is the justification of this method from the aspect of open systems.

When the Hamiltonian is Hermitian, the bath state is not influenced, and the normalizing factor $N_{\rho_{\rm tot}}(t)$ is always one. Therefore, the characteristic function of work in Eq.~(\ref{workchf}) becomes
\begin{eqnarray}
\chi(u)&=&\sum_{m,n}e^{iu(E_{m}(0)-E_{n}(0))}|\langle m({0})|U(t_{\rm f})\langle\psi_{n}^{\rm bath}|\Psi_{\rm tot}\rangle|^2\nonumber\\
       &=&\sum_{m,n}e^{iu(E_{m}(0)-E_{n}(0))}\frac{\exp(-\beta E_{n}(0))}{Z}\nonumber\\
       &&\times|\langle m({0})|U(t_{\rm f})|n(0)\rangle|^2,\nonumber\\
\end{eqnarray}
which is the ordinary expression for work statistics in a Hermitian system. After this simple situation, we consider general cases.

First, we formally retain the dynamics in the total Hilbert space shown in Fig.~{\ref{fig1}}(a), which is closed and Hermitian. The Hamiltonian and the state in the total Hilbert space are expressed as $H_{\rm THS}(t)$ and $|\psi_{\rm THS}(t)\rangle$, respectively. The state evolves according to the following relation,
\begin{eqnarray}
\frac{\partial}{\partial t}|\psi_{\rm THS}(t)\rangle&=&-iH_{\rm THS}(t)|\psi_{\rm THS}(t)\rangle.\label{THSequation}
\end{eqnarray}
We can introduce the projection operator onto the observed Hilbert space $P$, and the one onto the remaining part of the total Hilbert space $Q=I-P$. By applying $P$ and $Q$ in Eq.~(\ref{THSequation}), the following equation can be obtained:
\begin{eqnarray}
\frac{\partial}{\partial t}P|\psi_{\rm THS}(t)\rangle&=&-iPH_{\rm THS}(t)(P+Q)|\psi_{\rm THS}(t)\rangle\nonumber\\
                                                     &=&-i[PH_{\rm THS}(t)P]P|\psi_{\rm THS}(t)\rangle\nonumber\\
                                                     &&-iPH_{\rm THS}(t)Q|\psi_{\rm THS}(t)\rangle
\end{eqnarray}
If we design a proper $H_{\rm THS}(t)$, it is possible to satisfy the relation
\begin{eqnarray}
-iPH_{\rm THS}(t)Q|\psi_{\rm THS}(t)\rangle&=&-iM(t)P|\psi_{\rm THS}(t)\rangle,\nonumber\\\label{constrainth}
\end{eqnarray}
where $M$ is a non-Hermitian matrix in the observed Hilbert space. With Eq.~(\ref{constrainth}), the non-Hermitian evolution can be expressed as a projection of the dynamics in a larger Hilbert space.
\begin{eqnarray}
\frac{\partial}{\partial t}|\Psi_{\rm tot}(t)\rangle&\equiv&\frac{\partial}{\partial t}P|\psi_{\rm THS}(t)\rangle\nonumber\\
                                          &=&-i[PH_{\rm THS}(t)P+M(t)]P|\psi_{\rm THS}(t)\rangle\nonumber\\
                                          &=&-i[H_{\rm nh}(t)+H_0]|\Psi_{\rm tot}(t)\rangle,\label{dyprojection}
\end{eqnarray}
with $|\Psi_{\rm tot}(t)\rangle\equiv P|\psi_{\rm THS}(t)\rangle$. Such a projection can usually be realized by introducing an ancillary system and measurements on the ancillary system~\cite{nhmeasurementge1,nhmeasurementge2,nhmeasurementge3}.

Second, we apply the method presented in the previous subsection to the total Hilbert space. As the non-Hermitian part $H_{\rm nh}$ is turned off at the beginning time and the ending time, the observed space and the remaining part are decoupled at these two time points. So we have the relation
\begin{eqnarray}
H_{\rm THS}&=&PH_{\rm THS}P+QH_{\rm THS}Q.
\end{eqnarray}
To better show the relation between the work statistics in the total Hilbert space and the one in the observed space, we divide the state $|\psi_{\rm THS}\rangle$ into two parts and introduce basis states defined in Eq.~({\ref{purification}}),
\begin{eqnarray}
|\psi_{\rm THS}\rangle&=&\sum_nC^P_n|n,P\rangle\otimes|\psi^{\rm bath}_{n,P}\rangle\nonumber\\
&&+\sum_nC^Q_n|n,Q\rangle\otimes|\psi^{\rm bath}_{n,Q}\rangle,
\end{eqnarray}
where,
\begin{eqnarray}
P|n,P\rangle\otimes|\psi^{\rm bath}_{n,P}\rangle&=&|n,P\rangle\otimes|\psi^{\rm bath}_{n,P}\rangle,\nonumber\\
Q|n,Q\rangle\otimes|\psi^{\rm bath}_{n,Q}\rangle&=&|n,Q\rangle\otimes|\psi^{\rm bath}_{n,Q}\rangle.\nonumber
\end{eqnarray}
The form of $|n,P\rangle$ is decided by $H_{0}$, but the form of the other states can have many possibilities as long as they satisfy Eqs.~(\ref{purification}) and (\ref{bathbasis}). Note that the observed space is formed by the basis $|n,P\rangle\otimes|\psi^{\rm bath}_{m,P}\rangle$, which is the same as that formed by $|n(0)\rangle\otimes|\psi^{\rm bath}_{m}(t_{\rm f})\rangle$ in Eq.~(\ref{nhworkstatistics:finalbasis}). Therefore, we also have
\begin{eqnarray}\label{relation:definitionofprojection}
P|n,P\rangle\otimes|\psi^{\rm bath}_{m,Q}\rangle&=&0,\nonumber\\
P|n,Q\rangle\otimes|\psi^{\rm bath}_{m,P}\rangle&=&0.
\end{eqnarray}
The characteristic function of work in the total Hilbert space can then be divided into four parts:
\begin{eqnarray}
\chi_{\rm THS}(u)&=&\sum_{m,n}\left[e^{iu(E_{m,P}-E_{n,P})}P^{PP}_{m,n}+e^{iu(E_{m,P}-E_{n,Q})}P^{QP}_{m,n}\right.\nonumber\\
       &&+\left.e^{iu(E_{m,Q}-E_{n,Q})}P^{QQ}_{m,n}+e^{iu(E_{m,Q}-E_{n,Q})}P^{PQ}_{m,n}\right],\nonumber
\end{eqnarray}
with
\begin{eqnarray}\label{relation:totalspacestatistics}
&&H_{\rm THS}|n,P\rangle=E_{n,P}|n,P\rangle,\nonumber\\
&&H_{\rm THS}|n,Q\rangle=E_{n,Q}|n,Q\rangle,\nonumber\\
&&P^{PP}_{m,n}=\langle m,P|\otimes\langle\psi^{\rm bath}_{n,P}(t_{\rm f})|\rho_{\rm THS}(t_{\rm f})|m,P\rangle\otimes|\psi^{\rm bath}_{n,P}(t_{\rm f})\rangle,\nonumber\\
&&P^{QP}_{m,n}=\langle m,P|\otimes\langle\psi^{\rm bath}_{n,Q}(t_{\rm f})|\rho_{\rm THS}(t_{\rm f})|m,P\rangle\otimes|\psi^{\rm bath}_{n,Q}(t_{\rm f})\rangle,\nonumber\\
&&P^{QQ}_{m,n}=\langle m,Q|\otimes\langle\psi^{\rm bath}_{n,Q}(t_{\rm f})|\rho_{\rm THS}(t_{\rm f})|m,Q\rangle\otimes|\psi^{\rm bath}_{n,Q}(t_{\rm f})\rangle,\nonumber\\
&&P^{PQ}_{m,n}=\langle m,Q|\otimes\langle\psi^{\rm bath}_{n,P}(t_{\rm f})|\rho_{\rm THS}(t_{\rm f})|m,Q\rangle\otimes|\psi^{\rm bath}_{n,P}(t_{\rm f})\rangle.\nonumber\\
\end{eqnarray}

Here, the term $\rho_{\rm THS}(t_{\rm f})\equiv|\psi_{\rm THS}(t_{\rm f})\rangle\langle\psi_{\rm THS}(t_{\rm f})|$ is the density matrix at $t_{\rm f}$ in the total Hilbert space. The first part $\sum_{m,n}e^{iu(E_{m,P}-E_{n,P})}P^{PP}_{m,n}$ corresponds to the trajectories ending in the observed Hilbert space. Other terms describe the trajectories falling outside the observed Hilbert space.

As we have the relation in Eq.~(\ref{dyprojection}), it is straightforward to show that the characteristic function of work in the observed space is
\begin{eqnarray}\label{relationtoHermitian:partoftrajectories}
\chi(u)&=&\frac{N_{\rho_{\rm tot}}(0)}{N_{\rho_{\rm tot}}(t_{\rm f})}\sum_{m,n}e^{iu(E_{m,P}-E_{n,P})}P^{PP}_{m,n}.
\end{eqnarray}
Note that the Hamiltonian in the total Hilbert space $H_{\rm THS}(t)$ is Hermitian, so that our method just provides the results of the ordinary two-point measurement method. Therefore, the work statistics introduced by us in a non-Hermitian process is a statistics on part of the trajectories in the total Hilbert space [black and blue lines in Fig.~\ref{fig1}(a)]. Note that it is also possible to define $P|n,P\rangle\otimes|\psi^{\rm bath}_{m,Q}\rangle=|n,P\rangle\otimes|\psi^{\rm bath}_{m,Q}\rangle$ in Eq.~(\ref{relation:definitionofprojection}), which results in a different choice of the observed space. In this case, the basis states of the bath $|\psi_{n}^{\rm bath}\rangle$ in Eq.~(\ref{nhworkstatistics:finalbasis}) contain both $|\psi^{\rm bath}_{m,P}\rangle$ and $|\psi^{\rm bath}_{m,Q}\rangle$; the term $\sum_{m,n}e^{iu(E_{m,P}-E_{n,Q})}P^{QP}_{m,n}$ in Eq.~(\ref{relation:totalspacestatistics}) also contributes to the work statistics.

The relation in Eq.~(\ref{relationtoHermitian:partoftrajectories}) can also be useful when we want to calculate work statistics of non-Hermitian systems. Estimating Eq.~(\ref{workchf}) is usually difficult due to measurements on the bath. However, Eqs.~(\ref{relation:totalspacestatistics}) and (\ref{relationtoHermitian:partoftrajectories}) allow to obtain the non-Hermitian work statistics from the Hermitian work statistics in the total Hilbert space. We can first estimate the work statistics in the total Hilbert space by using the ordinary two-point measurement method. Next, we pick up the trajectories ending in the observed space and obtain the non-Hermitian work statistics. Note that if there is no energy exchange between system and bath in the total Hilbert space, measurements on the bath can be avoided.
\begin{figure}[t]
\center
\includegraphics[width=3.4in]{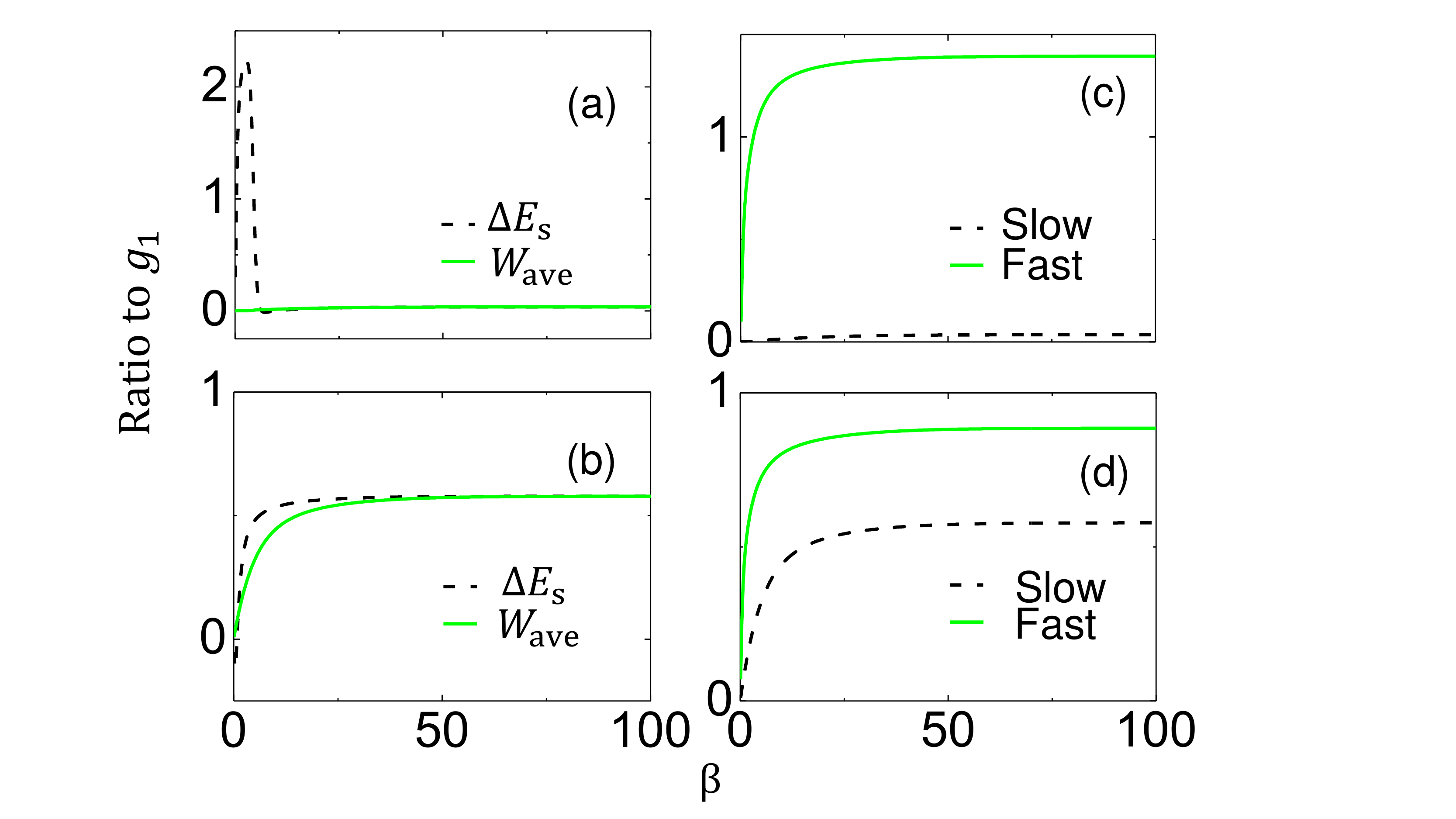}
\caption{(a) The average work $W_{\rm ave}$ and the system-energy change probability $\Delta E_s$ for $\gamma=1.9g_1$ (before the EP). (b) The average work $W_{\rm ave}$ and the system-energy change $\Delta E_s$ for $\gamma=2.1g_1$ (after the EP). (c) The comparison between the average works $W_{\rm ave}$ of a slowly changing Hamiltonian and a fast changing Hamiltonian for $\gamma=1.9g_1$ (before the EP). (d) The comparison between the average works $W_{\rm ave}$ of a slowly changing Hamiltonian and a fast changing Hamiltonian for $\gamma=2.1g_1$ (after the EP).
The inverse of the temperature $\beta$ and the energy are expressed in units of $1/g_1$ and $g_1$, respectively.}
\label{fig2}
\end{figure}
\section{Work statistics in the Non-Hermitian SSH model}
\label{workstatiscs}
As an example, we introduce a non-Hermitian SSH model based on the Hermitian version of the model~\cite{SSHmodel1,SSHmodel2,SSHmodel3,SSHmodel4}, which describes a one-dimensional topological chain with two different hopping integrals [Fig.~\ref{fig1}(b)]. The Hamiltonian of the Hermitian SSH model is
\begin{eqnarray}
H_0=\sum_n(g_1c^{\dag}_{2n}c_{2n+1}+g_2c^{\dag}_{2n+1}c_{2n+2}+{\rm h.c.}),
\label{SSHhamiltonian}
\end{eqnarray}
where $c_n$ and $c_n^{\dag}$ are the fermionic annihilation and creation operators on the $n$th site, respectively. Generally, the inner-unit-cell hopping integral $g_1$ and the inter-unit-cell hopping integral $g_2$ are not identical. In our examples, we assume $g_2=1.5g_1$. In addition to $H_0$ in Eq.~(\ref{SSHhamiltonian}), the non-Hermitian version of the model usually contains one of the following Hamiltonians:
\begin{eqnarray}
H^{\rm nr}_{\rm nh}(t)&=&\sum_nf(t)(\frac{\gamma}{2}c^{\dag}_{2n}c_{2n+1}-\frac{\gamma}{2}c^{\dag}_{2n+1}c_{2n}),\nonumber\\
H^{\rm lg}_{\rm nh}(t)&=&if(t)\delta\sum_n(c^{\dag}_{2n}c_{2n}-c^{\dag}_{2n+1}c_{2n+1}),
\end{eqnarray}
where the control function $f(t)$ is to turn on or off the non-Hermitian Hamiltonians. The first Hamiltonian $H^{\rm nr}_{\rm nh}(t)$ makes the hopping integrals for the left and the right directions unequal and it is usually called the nonreciprocal Hamiltonian. The second Hamiltonian $H^{\rm lg}_{\rm nh}(t)$ introduces gain and loss to the system, which is usually thought to be semi-classical. Both non-Hermitian SSH models have PT symmetry and their eigenstates can break this symmetry by passing through the exceptional point (EP). These models can be realized in, e.g., optical systems or cold atoms~\cite{genhsys1,genhsys2,nhtopo3,nhtopo4,nhSSHer1,nhSSHer3,nhSSHer4,nhSSHer5}, but the experimental demonstration of the effects in the quantum regime is still facing a big challenge. In the following two subsections, we consider the work statistics corresponding to these two kinds of non-Hermitian terms.

\subsection{Nonreciprocal coupling}
Before studying more complicated work statistics, we first consider simple quantities such as the average work and the change of the system's energy to reveal some unique properties of non-Hermitian systems. The non-Hermitian Hamiltonian $H_{\rm nh}^{\rm nr}(t)$ is introduced by using a {\it slowly} changing control function $f(t)=\sin(\pi t/T_{\rm tot})$, where $T_{\rm tot}=500/g_1$ is the total evolution time. The results are shown in Figs.~\ref{fig2}(a) and \ref{fig2}(b) for the PT-symmetric and PT-symmetry-breaking phases, respectively. First, we compare the average work with the system-energy change. The work is not equal to the system-energy change except for the regime of very low temperatures. Although the system and the bath are not directly coupled, the norm non-conserving property of the non-Hermitian Hamiltonian can influence the bath via the initial system-bath entanglement. Consequently, it is difficult to define the work without considering the bath.

Then, we focus on the average work in different cases. It is not surprising to see a vanishing amount of work in the PT-symmetric phase because the evolution is nearly {\it adiabatic}. If the system enters the regime of PT-symmetry-breaking phase during the process, the amount of the work increases significantly. It is insightful to discuss {\it non-adiabatic} effects. We consider a fast changing control function: $f(t)=1$, $t\in [0,T_{\rm tot}]$, and $f=0$ otherwise; namely, the non-Hermitian Hamiltonian is turned on and off suddenly. Figures~\ref{fig2}(c) and \ref{fig2}(d) show the work corresponding to both slow and fast changes. Similar to the Hermitian case, the sudden change causes non-adiabatic effects and increases the work. However, the non-adiabatic effects are significantly suppressed after passing through the EP.

\begin{figure}[t]
\center
\includegraphics[width=3.4in]{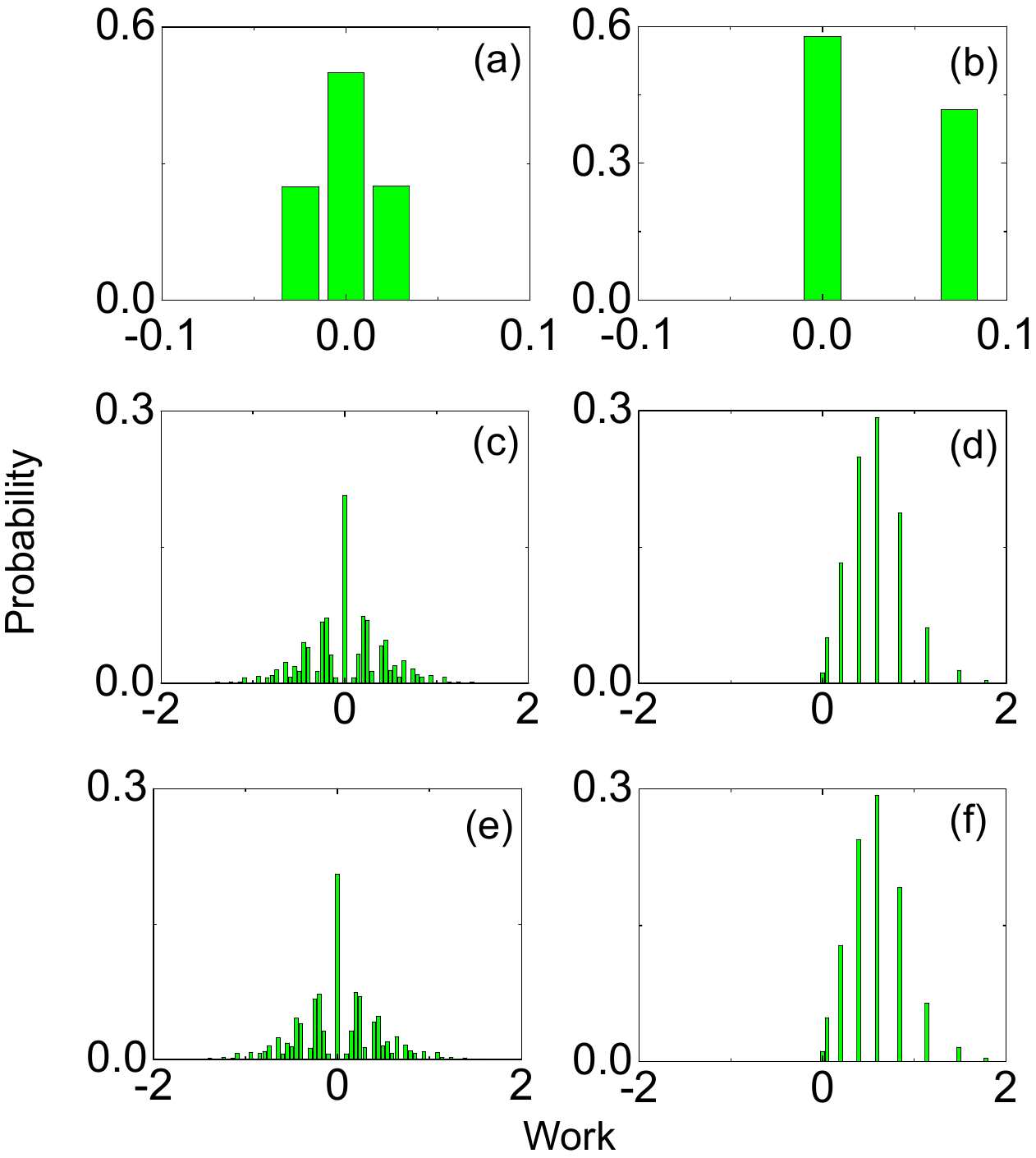}
\caption{(a) High temperature case $\beta=0.1/g_1$ with $\gamma=1.9$ (before the EP). (b) Low temperature case $\beta=1000/g_1$ with $\gamma=1.9$ (before the EP). (c) High temperature case $\beta=0.1/g_1$ with $\gamma=2.1$ (after the EP). (d) Low temperature case $\beta=1000/g_1$ with $\gamma=2.1$ (after the EP). (e) High temperature case $\beta=0.1/g_1$ with $\gamma=2.1$ (after the EP) and one additional round. (f) Low temperature case $\beta=1000/g_1$ with $\gamma=2.1$ (after the EP) and one additional round.
The inverse of the temperature $\beta$ and the energy are expressed in units of $1/g_1$ and $g_1$, respectively. The change of the Hamiltonian is slow.}
\label{fig3}
\end{figure}
\begin{figure*}[t]
\center
\includegraphics[width=7.2in]{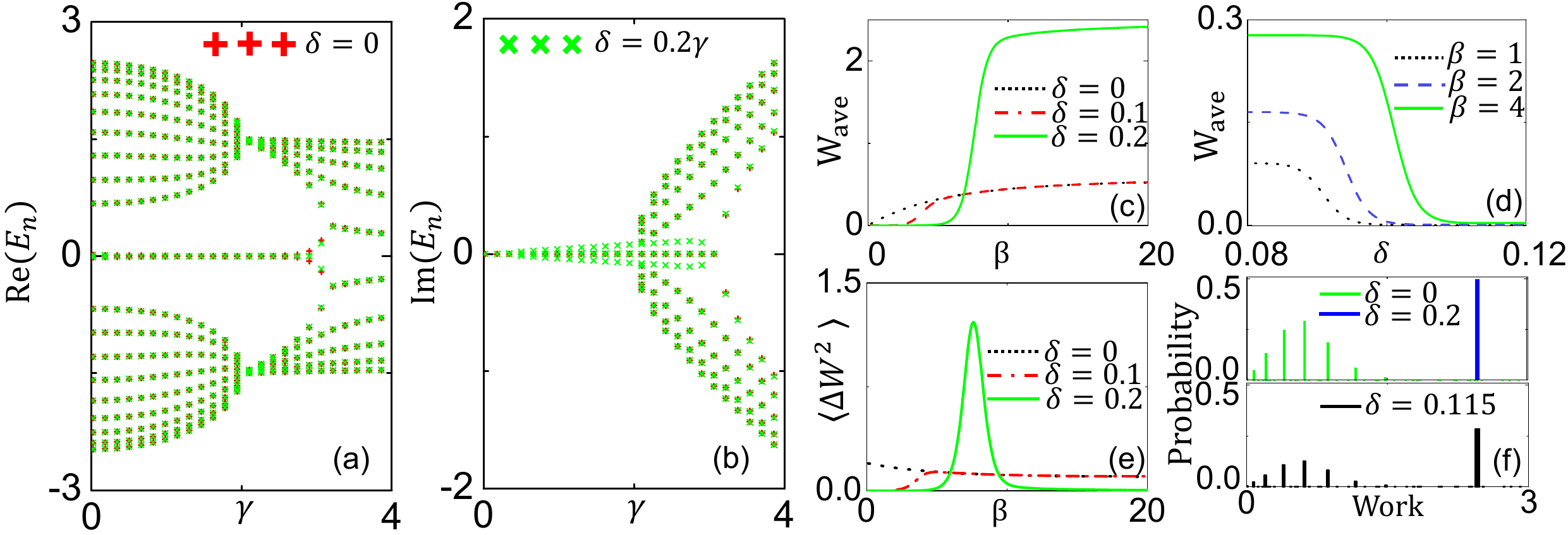}
\caption{The effects of the additional non-Hermitian Hamiltonian. (a) The real parts of the eigenvalues for systems with $\delta=0$ and $\delta=0.2\gamma$, respectively. (b) The imaginary parts of the eigenvalues for systems with $\delta=0$ and $\delta=0.2\gamma$, respectively. (c) The average work $W_{\rm ave}$ versus $\beta$ with $\gamma=2.1g_1$ (after the EP) and different values of $\delta$. (d) The average work $W_{\rm ave}$ versus $\delta$ with $\gamma=2.1g_1$ (after the EP) and different values of $\beta$. (e) The fluctuation of the work $\langle\Delta W^2\rangle$ versus $\beta$ with $\delta=0,~0.1\gamma$, and $0.2\gamma$.  (f) The work probabilities with $\beta=1000/g_1$ and different values of $\delta$.
The work and $\gamma$ are expressed in units of $g_1$, and $\delta$ is shown as the ratio to $\gamma$. The inverse of the temperature $\beta$ is expressed in units of $1/g_1$. The control function is set to be $f(t)=1$ in (a) and (b). The strength $\gamma$ of $H_{\rm nh}^1$ is $2.1g_1$ (after the EP) in (c)-(f).}
\label{fig4}
\end{figure*}
In addition to the average values, fluctuations can also play an important role in thermodynamics. Therefore, we consider the fluctuation of the work corresponding to the generation of a non-Hermitian SSH model. The statistics of the work is expressed with the work distribution,
\begin{eqnarray}
P(w)&=&\sum_{m,n}\delta(w-E_m+E_n)P_{m,n},
\end{eqnarray}
which is shown in Fig.~\ref{fig3}. From the numerical results, we can find that the fluctuation of the work increases significantly after passing through the EP. The average work in Fig.~(\ref{fig2}) always vanishes at high temperatures in spite of the value of $\gamma$, but the fluctuation of the work can change considerably for different values of $\gamma$. The fluctuations in Fig.~\ref{fig3}(c) are much larger than those in Fig.~\ref{fig3}(a). As a result, it is still possible to detect a large amount of work in spite of the vanishing average value after passing through the EP. This distribution becomes different when the temperature is low. In Fig.~\ref{fig3}(d), the distribution of the work is always positive and concentrated on several values.
We have mentioned that the slow evolution can be somewhat ``adiabatic''. This statement can be partially confirmed by the results in Figs.~\ref{fig3}(a) and (b). The work distribution $P_{m,n}$ is also the distribution of the transition among different states. Therefore, the transition among most eigenstates are negligible, and the change is nearly adiabatic. However, after passing the EP, such ``adiabatic'' condition does not exist anymore as shown in Figs.~\ref{fig3}(c) and (d).

Although there are no adiabatic like properties after passing EP, the system may enter a steady state (or metastable state~\cite{metastable}) after the long-time evolution. To check this point, we add one more round of evolution, namely, repeat the time-dependent non-Hermitian Hamiltonian during $t\in[0,T_{\rm tot}]$. After two rounds of evolution, we calculate the work statistics and show the results in Figs.~\ref{fig3}(e) and (d). The results of two-round evolution are nearly the same as those results of one-round evolution. Therefore, we still have metastable behavior after passing EP.
\subsection{Adding loss-gain terms}
Now, we consider a more complicated case, in which the non-Hermitian Hamiltonian $H_{\rm nh}^{\rm lg}$ is further introduced to $H_0+H_{\rm nh}^{\rm nr}$. To show the influence of this term, we study the slow evolutions in Figs.~\ref{fig2} and \ref{fig3} with the system described by $H_0+H_{\rm nh}^{\rm nr}+H_{\rm nh}^{\rm lg}$.

The properties of non-Hermitian systems are usually expressed with their eigenvalue spectra. Therefore, we first consider the influence of the additional term $H_{\rm nh}^{\rm lg}$ on the spectrum of the system. To have a clear picture of this effect, we compare the eigenvalues $E_n$ of the system after including $H_{\rm nh}^{\rm lg}$ ($\delta=0.2\gamma$) with those of the original system $H_0+H_{\rm nh}^{\rm nr}$ in Figs.~\ref{fig4}(a) and \ref{fig4}(b), where we choose $f(t)=1$. The results show that both the real and imaginary parts of $E_n$ are only slightly changed. However, the imaginary parts become nonzero before approaching the EP of the system {\it without} $H_{\rm nh}^{\rm lg}$. Note that only eigenstates with the smallest imaginary parts survive in the slow processes considered here, so small changes of the imaginary parts can have a significant influence on the system evolution.

Now, we consider the effects of the additional term on work statistics. In Fig.~\ref{fig4}(c) we study the change of the average work done at different temperatures. Compared with the results in Fig.~\ref{fig2}(b), the average work is reduced at moderate temperatures but increased at low temperatures. In addition, a large strength $\delta$ can suppress $W_{\rm ave}$ in a wider range of temperatures but the negative effects at low temperatures are more significant. In Fig.~\ref{fig4}(d), the average work only changes with $\delta$ in a very small regime.  The fluctuations of work are shown with the quantity $\langle\Delta W^2\rangle\equiv\sum_{\omega}P(\omega)(\omega-W_{\rm ave})^2$ in Figs.~\ref{fig4}(e). The additional term can suppress the fluctuations at most temperatures but can have the opposite function at some temperatures. The fluctuation increases near the temperature at which the average work in Fig.~\ref{fig4}(c) begins to rise. However, unlike the average work, the fluctuation vanishes if the temperature further decreases.

To understand the mechanism of $H_{\rm nh}^{\rm lg}$ more clearly, we present the work distribution with different values of $\delta$ in Fig.~\ref{fig4}(f). The green (or light gray upper) and blue (or dark gray upper) bars correspond to the cases without any control and with a strong enough control, respectively. It is clearly shown that these two distributions are totally different. When the intensity $\delta$ is not very strong (the black lower bars), the distribution of the work is the superposition of the green (or light gray upper) and the blue (or dark gray upper) ones. We can see from these results that the function of $H_{\rm nh}^{\rm lg}$ is to switch the system to another {metastable state} during the evolution process. This new metastable state has a different work statistics. Therefore, two non-Hermitian systems with similar spectra can have totally different thermodynamical properties in a process.

\section{Conclusions}
\label{conclusion}
We have studied the work statistics of a temporal non-Hermitian model based on Hermitian one, and take the non-Hermitian SSH model as an example. To obtain the work statistics for such a non-Hermitian process, the ordinary work estimation approach, which does not work for the non-Hermitian cases, is modified here. Our method can be applied in PT broken cases, and becomes the ordinary two-point measurement method in Hermitian systems. With this method, we analyze the work statistics in the non-Hermitian SSH model. After passing through the EP, the work statistics changes considerably. The average work becomes larger at low temperatures, and the work fluctuations significantly increase at all temperatures. We further consider another non-Hermitian Hamiltonian with similar spectrum. In spite of the negligible spectra change, the work fluctuations are suppressed at most temperatures, and the average work is also reduced at moderate temperatures. Therefore, the work statistics is able to provide system information which cannot be clearly presented by the spectrum.

Our work partially solves the problem of work statistics in non-Hermitian systems, especially in PT broken cases. This method can be applied to many non-Hermitian problems, and can be potentially extended to more general situations.

\section{Acknowledgments}
We are grateful to Dr.~Ken Funo and Dr.~Tao Liu for their instructive comments. Z.L.X. is supported by the National Natural Science
Foundation of China (NSFC) (Grants No.~11874432). J.Q.Y. is supported by the National Natural Science Foundation of China (NSFC) (Grant No. 11774022 and No. U1801661). F.N. is supported in part by: Nippon Telegraph and Telephone Corporation (NTT) Research, the Japan Science and Technology Agency (JST)[via the Quantum Leap Flagship Program (Q-LEAP), the Moonshot R{\&}D Grant Number JPMJMS2061, and the Centers of Research Excellence in Science and Technology (CREST) Grant No. JPMJCR1676], the Japan Society for the Promotion of Science (JSPS) [via the Grants-in-Aid for Scientific Research (KAKENHI) Grant No. JP20H00134 and the JSPS-RFBR Grant No. JPJSBP120194828], the Army Research Office (ARO) (Grant No. W911NF-18-1-0358), the Asian Office of Aerospace Research and Development (AOARD) (via Grant No. FA2386-20-1-4069), and the Foundational Questions Institute Fund (FQXi) via Grant No. FQXi-IAF19-06.

\appendix
\section{Comparison between normalization and bi-orthogonal basis}
In a non-Hermitian evolution, the ordinary state norm is not conserved
\begin{eqnarray}
\frac{\partial}{\partial t}\langle \psi(t)|\psi(t)\rangle&=&i\langle \psi(t)|H^{\dag}|\psi(t)\rangle-i\langle \psi(t)|H|\psi(t)\rangle\nonumber\\
&\neq&0.
\end{eqnarray}
There are two main methods to deal with this problem.

One way is to normalize the state:
\begin{eqnarray}
|\tilde{\psi}(t)\rangle&=&\frac{|\psi(t)\rangle}{\sqrt{\langle \psi(t)|\psi(t)\rangle}}.
\end{eqnarray}
The density matrices can also be normalized in the same way~\cite{nhevo},
\begin{eqnarray}
\tilde{\rho}(t)&=&\frac{\rho(t)}{{\rm Tr}\{\rho(t)\}},
\end{eqnarray}
where $\rho(t)$ is a density matrix under the influence of a non-Hermitian Hamiltonian.

This method has several advantages.

First, all the quantities, e.g., state, density matrix, or expectation values, have the same properties as those in the Hermitian case.

Second, this framework is not influenced by the parity-time symmetry. However, effects of parity-time symmetry breaking can still be captured with this method.

Third, this method can be interpreted as the process of post-selection. When a normalized state $|\psi_{\rm N}\rangle$ is measured, we can pick up data corresponding to a subspace, e.g., lower two levels.

The observed state $|\psi\rangle$ is in general not normalized as
\begin{eqnarray}
&&|\psi\rangle=P|\psi_{\rm N}\rangle,\nonumber\\
&&\langle\psi|\psi\rangle+\langle\psi_{\rm N}|(I-P)|\psi_{\rm N}\rangle=1.
\end{eqnarray}
Here, $P$ is a projection operator and $I$ is the identity operator. We call this kind of non-Hermitian systems ``open" due to the existence of an additional space.

However, this kind of normalization can significantly change the dynamical and thermodynamic properties of the system. Thermodynamic relations may be broken. In addition, it is usually impossible to find the eigenstates and eigenvalues in this framework.

Another way to deal with non-Hermitian systems is to introduce a bi-orthogonal basis (eigenstates)~\cite{biorthermo1,biorthermo2,nhevo2}.
\begin{eqnarray}
H|\psi_n\rangle&=&E_n|\psi_n\rangle,\nonumber\\
H^{\dag}|\phi_n\rangle&=&E^*_n|\phi_n\rangle,\nonumber\\
\langle\phi_n|\psi_m\rangle&=&\delta_{m,n}.
\end{eqnarray}
This method can obtain the eigenvalues and eigenstates of a non-Hermitian Hamiltonian, and is very useful in the parity-time symmetry phase. When the system has parity-time symmetry, the bi-orthogonal basis becomes an orthogonal basis with a non-trivial metric $M$,
\begin{eqnarray}
H|\psi_n\rangle&=&E_n|\psi_n\rangle,\nonumber\\
H^{\dag}M|\psi_n\rangle&=&E_nM|\psi_n\rangle,\nonumber\\
\langle\psi_n|M|\psi_m\rangle&=&\delta_{m,n}.
\end{eqnarray}
In addition, the non-Hermitian dynamics can be mapped onto a Hermitian dynamics~\cite{biorthermo3}. Therefore, all the thermodynamic relations in Hermitian systems should conceptually have non-Hermitian counterparts.

However, this method is less powerful in parity-time symmetry broken cases. Inner products (including state norms) are not conserved; thermodynamic relations are no longer assured, and, complex eigenvalues cannot be simply interpreted as energies.
\section{Description of the heat bath}
In this appendix, we briefly discuss the definition of the heat bath used in this manuscript. Assume that there is a large system B in thermal equilibrium,
\begin{eqnarray}
\rho_{\rm B}&=&\frac{\sum_ne^{-\beta \tilde{E}_n}\tilde{|n\rangle}\tilde{\langle n|}}{\sum_ne^{-\beta {\tilde E}_n}}.
\end{eqnarray}
Here, $\tilde{|n\rangle}$ and $\tilde{E}_n$ are eigenstates and eigenvalues, respectively. System B can be used as a heat bath to drive a small system A into a thermal equilibrium state.
\begin{eqnarray}
\rho_{\rm A}&=&{\rm Tr_{B}}\left\{U(\infty)|\psi_{\rm A}\rangle\langle\psi_{\rm A}|\otimes\rho_{\rm B}U^{\dag}(\infty)\right\}\nonumber\\
            &\approx&\frac{\sum_ne^{-\beta E_n}|n\rangle\langle n|}{\sum_ne^{-\beta E_n}}.
\end{eqnarray}
The eigenstates and eigenvalues of system A are described by $|n\rangle$ and $E_n$, respectively. The coupling between system A and system B is included by the evolution operator $U$. Next, we show that this A+B model can be derived from the form of Eq.~(\ref{purification}).

Note that the reduced density matrix $\rho_{\rm A}$ is conceptually an average of a pure state over some degrees of freedom,
\begin{eqnarray}
\rho_{\rm A}&=&{\rm Tr}_{\rm other\ than\ A}\{|\psi_{\rm world}\rangle\langle\psi_{\rm world}|\}.
\end{eqnarray}
The pure state $|\psi_{\rm world}\rangle$ is usually unknown but can be simplified by applying the eigenstate thermalization hypothesis,
\begin{eqnarray}
|\psi_{\rm world}\rangle&=&|\Psi_{\rm tot}\rangle\otimes|\psi_{\rm other}\rangle.
\end{eqnarray}
The state $|\Psi_{\rm tot}\rangle$ is defined in Eq. (\ref{purification}); the state $|\psi_{\rm other}\rangle$ has no effects on the thermodynamics properties of system A. Therefore, we have the relation
\begin{eqnarray}
\rho_{\rm A}&=&{\rm Tr}_{\rm other\ than\ A}\{|\Psi_{\rm tot}\rangle\langle\Psi_{\rm tot}|\}.
\end{eqnarray}
Note that $|\Psi_{\rm tot}\rangle$ covers the Hilbert space of system B because system A and system B become entangled during the coupling. In addition, the heat bath system B is usually assumed to be unchanged during the coupling with system A, so that we also have
\begin{eqnarray}
\rho_{\rm B}&=&{\rm Tr}_{\rm other\ than\ B}\{|\Psi_{\rm tot}\rangle\langle\Psi_{\rm tot}|\}.
\end{eqnarray}
If we describe the thermalization of system A with the coupling between $|\psi_{\rm A}\rangle\langle\psi_{\rm A}|$ and $\rho_{\rm B}$, the effects of the heat bath can be characterized by a classical distribution.

Such a description becomes insufficient when post-selection is applied to generate non-Hermitian dynamics,
\begin{eqnarray}
\tilde{\rho_{\rm A}}(t)&=&P\rho_{\rm A}(t)P.
\end{eqnarray}
Here, $P$ is a projection operator in the Hilbert space of system A. This post-selection can influence the Hilbert space of system B as follows,
\begin{eqnarray}
\rho_{\rm B}\neq{\rm Tr}_{\rm other\ than\ B}\{P|\Psi_{\rm tot}\rangle\langle\Psi_{\rm tot}|P\}.
\end{eqnarray}
Therefore, we use $|\psi_{m}^{\rm bath}\rangle$ instead of a classical distribution $\rho_{\rm B}$ to describe the bath effects in non-Hermitian processes.



%


\begin{references}
%
\bibitem{genhsys1}R. El-Ganainy, K. G. Makris, D. N. Christodoulides, and Z. H. Musslimani, Theory of coupled optical PT-symmetric structures, \href{https://doi.org/10.1364/OL.32.002632}{Opt. Lett. {\bf 32}, 2632 (2007)}.
%
%
\bibitem{genhsys2}A. Guo, G. J. Salamo, D. Duchesne, R. Morandotti, M. Volatier-Ravat, V. Aimez, G. A. Siviloglou, and D. N. Christodoulides,Observation of PT-symmetry breaking in complex optical potentials,  \href{https://doi.org/10.1103/PhysRevLett.103.093902}{Phys. Rev. Lett. {\bf 103}, 093902 (2009)}.
%
%
\bibitem{nhsys2}C. E. R$\rm \ddot{u}$ter, K. G. Makris, R. El-Ganainy, D. N. Christodoulides, M. Segev, and D. Kip, Observation of parity-time symmetry in optics, \href{https://doi.org/10.1038/NPHYS1515}{Nat. Phys. {\bf 6}, 192 (2010)}.
%
%
\bibitem{nhsys3}L. Feng, Z. J. Wong, R.-M. Ma, Y. Wang, X. Zhang, Single-mode laser by parity-time symmetry breaking, \href{https://doi.org/10.1126/science.1258479}{Science {\bf 346}, 972 (2014)}.
%
%
\bibitem{nhsys4}B. Peng, S. K. ${\rm \ddot{O}}$zdemir, F. Lei, F. Monifi, M. Gianfreda, G. L. Long, S. Fan, F. Nori, C. M. Bender, and L. Yang, Parity-time-symmetric whispering-gallery
microcavities, \href{https://doi.org/10.1038/NPHYS2927}{Nat. Phys. {\bf 10}, 294 (2014)}.
%
%
\bibitem{nhsys5}T. Gao, E. Estrecho, K. Y. Bliokh, T. C. H. Liew, M. D. Fraser, S. Brodbeck, M. Kamp, C. Schneider, S. H$\rm \ddot{o}$fling, Y. Yamamoto, F. Nori, Y. S. Kivshar, A. G. Truscott, R. G. Dall, and E. A. Ostrovskaya, Observation of non-Hermitian degeneracies in a chaotic exciton-polariton billiard, \href{https://doi.org/10.1038/nature15522}{Nature {\bf 526}, 554 (2015)}.
%
%
\bibitem{nhsys6}J. M. Zeuner, M. C. Rechtsman, Y. Plotnik, Y. Lumer, S. Nolte, M. S. Rudner, M. Segev, and A. Szameit, Observation of a topological transition in the bulk of a non-Hermitian system, \href{https://doi.org/10.1103/PhysRevLett.115.040402}{Phys. Rev. Lett. {\bf 115}, 040402 (2015)}.
%
%
\bibitem{nhsys1}R. El-Ganainy, K. G. Makris, M. Khajavikhan, Z. H. Musslimani, S. Rotter, and D. N. Christodoulides, Non-Hermitian physics and PT symmetry, \href{https://doi.org/10.1038/NPHYS4323}{Nat. Phys. {\bf 14}, 11 (2018)}.
%
%
\bibitem{nhsys7}S. K. ${\rm \ddot{O}}$zdemir, S. Rotter, F. Nori, and L. Yang, Parity-time symmetry and exceptional points in photonics, \href{https://doi.org/10.1038/s41563-019-0304-9}{Nat. Mater. {\bf 18}, 783 (2019)}.
%
%
\bibitem{PTbook1}D. Christodoulides, J. Yang, Parity-time symmetry and its applications, Springer (Singapore) (2018).
%
%
\bibitem{nhtopo1}W. Hu, H. Wang, P. P. Shum, and Y. D. Chong, Exceptional points in a non-Hermitian topological pump, \href{https://doi.org/10.1103/PhysRevB.95.184306}{Phys. Rev. B {\bf 95}, 184306 (2017)}.
%
%
\bibitem{nhtopo2}D. Leykam, K. Y. Bliokh, C. Huang, Y. D. Chong, and F. Nori, Edge modes, degeneracies, and topological numbers in non-Hermitian systems, \href{https://doi.org/10.1103/PhysRevLett.118.040401}{Phys. Rev. Lett. {\bf 118}, 040401 (2017)}.
%
%
\bibitem{nhtopo3}Z. Gong, Y. Ashida, K. Kawabata, K. Takasan, S. Higashikawa, and M. Ueda, Topological phases of non-Hermitian systems, \href{https://doi.org/10.1103/PhysRevX.8.031079}{Phys. Rev. X {\bf 8}, 031079 (2018)}.
%
%
\bibitem{nhtopo4}T. Liu, Y.-R. Zhang, Q. Ai, Z. Gong, K. Kawabata, M. Ueda, and F. Nori, Second-order topological phases in non-Hermitian systems, \href{https://doi.org/10.1103/PhysRevLett.122.076801}{Phys. Rev. Lett. {\bf 122}, 076801 (2019)}.
%
%
\bibitem{nhtopo5}K. Y. Bliokh, D. Leykam, M. Lein, and F. Nori, Topological non-Hermitian origin of surface
Maxwell waves, \href{https://doi.org/10.1038/s41467-019-08397-6}{Nat. Commun. {\bf 10}, 580 (2019)}.
%
%
\bibitem{nhmb1}T. E. Lee and C.-K. Chan, Heralded magnetism in non-Hermitian atomic systems, \href{https://doi.org/10.1103/PhysRevX.4.041001}{Phys. Rev. X {\bf 4}, 041001 (2014)}.
%
%
\bibitem{nhmb2}T. E. Lee, F. Reiter, and N. Moiseyev, Entanglement and spin squeezing in non-Hermitian phase transitions, \href{https://doi.org/10.1103/PhysRevLett.113.250401}{Phys. Rev. Lett. {\bf 113}, 250401 (2014)}.
%
%
\bibitem{nhap1}S. Ib$\rm\acute{a}\tilde{n}$ez, S. Mart$\rm \acute{i}$nez-Garaot, X. Chen, E. Torrontegui, and J. G. Muga, Shortcuts to adiabaticity for non-Hermitian systems, \href{https://doi.org/10.1103/PhysRevA.84.023415}{Phys. Rev. A {\bf 84}, 023415 (2011)}.
%
%
\bibitem{nhap2}B. T. Torosov, G. D. Valle, and S. Longhi, Non-Hermitian shortcut to adiabaticity, \href{https://doi.org/10.1103/PhysRevA.87.052502}{Phys. Rev. A {\bf 87}, 052502 (2013)}.
%
%
\bibitem{nhap3}S. Ib$\rm\acute{a}\tilde{n}$ez and J. G. Muga, Adiabaticity condition for non-Hermitian Hamiltonians, \href{https://doi.org/10.1103/PhysRevA.89.033403}{Phys. Rev. A {\bf 89}, 033403 (2014)}.
%
%
\bibitem{nhap4}B. T. Torosov, G. D. Valle, and S. Longhi, Non-Hermitian shortcut to stimulated Raman adiabatic passage, \href{https://doi.org/10.1103/PhysRevA.89.063412}{Phys. Rev. A {\bf 89}, 063412 (2014)}.
%
%
\bibitem{nhap5}Y.-H. Chen, Y. Xia, Q.-C. Wu, B.-H. Huang, and J. Song, Method for constructing shortcuts to adiabaticity by a substitute of counterdiabatic driving terms, \href{https://doi.org/10.1103/PhysRevA.93.052109}{Phys. Rev. A {\bf 93}, 052109 (2016)}.
%
%
\bibitem{nhap6}Q.-C. Wu, Y.-H. Chen, B.-H. Huang, Y. Xia, and J. Song, Reverse engineering of a nonlossy adiabatic Hamiltonian for non-Hermitian systems, \href{https://doi.org/10.1103/PhysRevA.94.053421}{Phys. Rev. A {\bf 94}, 053421 (2016)}.
%
%
\bibitem{nhnonre1}X. Q. Li, X. Z. Zhang, G. Zhang, and Z. Song, Asymmetric transmission through a flux-controlled non-Hermitian scattering center, \href{https://doi.org/10.1103/PhysRevA.91.032101}{Phys. Rev. A {\bf 91}, 032101 (2015)}.
%
%
\bibitem{nhnonre2}C. Li, L. Jin, and Z. Song, Non-Hermitian interferometer: Unidirectional amplification without distortion, \href{https://doi.org/10.1103/PhysRevA.95.022125}{Phys. Rev. A {\bf 95}, 022125 (2017)}.
%
%
\bibitem{nhnonre3}T. T. Koutserimpas and R. Fleury, Nonreciprocal gain in non-Hermitian time-Floquet systems, \href{https://doi.org/10.1103/PhysRevLett.120.087401}{Phys. Rev. Lett. {\bf 120}, 087401 (2018)}.
%
%
\bibitem{nhlocal1}J. Feinberg and A. Zee, Non-Hermitian localization and delocalization, \href{https://doi.org/10.1103/PhysRevE.59.6433}{Phys. Rev. E {\bf 59}, 6433 (1999)}.
%
%
\bibitem{nhlocal2}J. A. S. Louren$\rm {c}$o, R. L. Eneias, and R. G. Pereira, Kondo effect in a PT-symmetric non-Hermitian Hamiltonian, \href{https://doi.org/10.1103/PhysRevB.98.085126}{Phys. Rev. B {\bf 98}, 085126 (2018)}.
%
%
\bibitem{nhlocal3}P. Wang, L. Jin, and Z. Song, Non-Hermitian phase transition and eigenstate localization induced by asymmetric coupling, \href{https://doi.org/10.1103/PhysRevA.99.062112}{Phys. Rev. A {\bf 99}, 062112 (2019)}.
%
%
\bibitem{nhlocal4}K. L. Zhang, X. M. Yang, and Z. Song, Quantum transport in non-Hermitian impurity arrays, \href{https://doi.org/10.1103/PhysRevB.100.024305}{Phys. Rev. B {\bf 100}, 024305 (2019)}.
%
%
\bibitem{SSHmodel1}W. P. Su, J. R. Schrieffer, and A. J. Heeger, Soliton excitations in polyacetylene, \href{https://doi.org/10.1103/PhysRevB.22.2099}{Phys. Rev. B {\bf 22}, 2099 (1980)}.
%
%
\bibitem{SSHmodel2}B. Zhu, R. L$\rm \ddot{u}$, and S. Chen, $\mathcal{PT}$ symmetry in the non-Hermitian Su-Schrieffer-Heeger model with complex boundary potentials, \href{https://doi.org/10.1103/PhysRevA.89.062102}{Phys. Rev. A {\bf 89}, 062102 (2014)}.
%
%
\bibitem{SSHmodel3}C. Yin, H. Jiang, L. Li, R. L$\rm \ddot{u}$, and S. Chen, Geometrical meaning of winding number and its characterization of topological phases in one-dimensional chiral non-Hermitian systems, \href{https://doi.org/10.1103/PhysRevA.97.052115}{Phys. Rev. A {\bf 97}, 052115 (2018)}.
%
%
\bibitem{SSHmodel4}S. Yao and Z. Wang, Edge states and topological invariants of non-Hermitian systems, \href{https://doi.org/10.1103/PhysRevLett.121.086803}{Phys. Rev. Lett. {\bf 121}, 086803 (2018)}.
%
%
\bibitem{expnhSSH}S. Weidemann, M. Kremer, T. Helbig, T. Hofmann, A. Stegmaier, M. Greiter, R. Thomale, and A. Szameit, Topological funneling of light, \href{https://doi.org/10.1126/science.aaz8727}{Science 10.1126/science.aaz8727 (2020)}.
%
%
\bibitem{quanthermre1}R. J. Harris and G. M. Sch$\rm \ddot{u}$tz, Fluctuation theorems for stochastic
dynamics, \href{https://doi.org/10.1088/1742-5468/2007/07/P07020}{J. Stat. Mech. {\bf 2007}, 07020 (2007)}.
%
%
\bibitem{quanthermre2}M. Esposito, U. Harbola, and S. Mukamel, Nonequilibrium fluctuations, fluctuation theorems, and counting statistics in quantum systems, \href{https://doi.org/10.1103/RevModPhys.81.1665}{Rev. Mod. Phys. {\bf 81}, 1665 (2009)}.
%
%
\bibitem{quanthermre3}M. Campisi, P. H$\rm\ddot{a}$nggi, and P. Talkner, $Colloquium$: Quantum fluctuation relations: Foundations and applications, \href{https://doi.org/10.1103/RevModPhys.83.771}{Rev. Mod. Phys. {\bf 83}, 771 (2011)}.
%
%
\bibitem{quanthermre4}H. T. Quan, Y. D. Wang, Y.-X. Liu, C. P. Sun, and F. Nori, Maxwell's demon assisted thermodynamic cycle in superconducting quantum circuits, \href{https://doi.org/10.1103/PhysRevLett.97.180402}{Phys. Rev. Lett. {\bf 97}, 180402 (2006)}.
%
%
\bibitem{quanthermre5}H. T. Quan, Y. D. Wang, Y.-X. Liu, C. P. Sun, and F. Nori, Quantum thermodynamic cycles and quantum heat engines, \href{https://doi.org/10.1103/PhysRevE.76.031105}{Phys. Rev. E {\bf 76}, 031105 (2007)}.
%
%
\bibitem{jequal1}C. Jarzynski, Nonequilibrium equality for free energy differences, \href{https://doi.org/10.1103/PhysRevLett.78.2690}{Phys. Rev. Lett. {\bf 78}, 2690 (1997)}.
%
%
\bibitem{tpm1}J. Kurchan, A quantum fluctuation theorem, \href{https://arxiv.org/abs/cond-mat/0007360}{arXiv:cond-mat/0007360}.
%
%
\bibitem{tpm2}H. Tasaki, Jarzynski Relations for quantum systems and some applications, \href{https://arxiv.org/abs/cond-mat/0009244}{arXiv:cond-mat/0009244}.
%
%
\bibitem{tpm3}P. Talkner, E. Lutz, and P. H$\rm \ddot{a}$nggi, Fluctuation theorems: Work is not an observable, \href{https://doi.org/10.1103/PhysRevE.75.050102}{Phys. Rev. E {\bf 75}, 050102(R) (2007)}.
%
%
\bibitem{npmw1}A. E. Allahverdyan, Nonequilibrium quantum fluctuations of work, \href{https://doi.org/10.1103/PhysRevE.90.032137}{Phys. Rev. E {\bf 90}, 032137 (2014)}.
%
%
\bibitem{npmw2}P. Solinas and S. Gasparinetti, Full distribution of work done on a quantum system for arbitrary initial states, \href{https://doi.org/10.1103/PhysRevE.92.042150}{Phys. Rev. E {\bf 92}, 042150 (2015)}.
%
%
\bibitem{npmw3}A. Ortega, E. McKay, A M. Alhambra, and E. Martin-Martinez, Work distributions on quantum fields, \href{https://doi.org/10.1103/PhysRevLett.122.240604}{Phys. Rev. Lett. {\bf 122}, 240604 (2019)}.
%
%
\bibitem{wwco1}P. Talkner and P. H$\rm \ddot{a}$nggi, Aspects of quantum work, \href{https://doi.org/10.1103/PhysRevE.93.022131}{Phys. Rev. E {\bf 93}, 022131 (2016)}.
%
%
\bibitem{wwco2}M. Perarnau-Llobet, E. B$\rm \ddot{a}$umer, K. V. Hovhannisyan, M. Huber, and A. Acin, No-go theorem for the characterization of work fluctuations in coherent quantum systems, \href{https://doi.org/10.1103/PhysRevLett.118.070601}{Phys. Rev. Lett. {\bf 118}, 070601 (2017)}.
%
%
\bibitem{wopen}K. Funo and H.T. Quan, Path integral approach to quantum thermodynamics, \href{https://doi.org/10.1103/PhysRevLett.121.040602}{Phys. Rev. Lett. {\bf 121}, 040602 (2018)}.
%
%
\bibitem{biorthermo1}J. Gong, and Q.-H. Wang, Time-dependent PT-symmetric quantum mechanics, \href{http://dx.doi.org/10.1088/1751-8113/46/48/485302}{J. Phys. A: Math. Theor. {\bf 46}, 485302 (2013)}.
%
%
\bibitem{biorthermo2}S. Deffner and A. Saxena, Jarzynski Equality in PT-Symmetric Quantum Mechanics, \href{https://doi.org/10.1103/PhysRevLett.114.150601}{Phys. Rev. Lett. {\bf 114}, 150601 (2015)}.
%
%
\bibitem{biorthermo3}B.-B. Wei, Quantum work relations and response theory in parity-time-symmetric quantum systems, \href{https://doi.org/10.1103/PhysRevE.97.012114}{Phys. Rev. E {\bf 97}, 012114 (2018)}.
%

%
\bibitem{nhevo}D. C. Brody and E.-M. Graefe, Mixed-state evolution in the presence of gain and loss, \href{https://doi.org/10.1103/PhysRevLett.109.230405}{Phys. Rev. Lett. {\bf 109}, 230405 (2012)}.
%
%
\bibitem{nhevo2}Chia-Yi Ju, Adam Miranowicz, Guang-Yin Chen, and Franco Nori, Non-Hermitian Hamiltonians and no-go theorems in quantum information, \href{https://doi.org/10.1103/PhysRevA.100.062118}{Phys. Rev. A {\bf 100}, 062118 (2019)}.
%
%
\bibitem{wfexp1}D. M. Carberry, M. A. B. Baker, G. M. Wang, E. M. Sevick, and D. J. Evans, An optical trap experiment to demonstrate fluctuation theorems in viscoelastic media,  \href{https://doi.org/10.1088/1464-4258/9/8/S13}{J. Opt. A {\bf 9}, 204 (2007)}.
%
%
%
%
\bibitem{wfexp3}Y. Masuyama, K. Funo, Y. Murashita, A. Noguchi, S. Kono, Y. Tabuchi, R. Yamazaki, M. Ueda, and Y. Nakamura, Information-to-work conversion by Maxwell's demon in a superconducting circuit quantum electrodynamical system, \href{https://doi.org/10.1038/s41467-018-03686-y}{Nat. Commun. {\bf 9}, 1291 (2018)}.
%
%
\bibitem{nhwec}A. Sergi and K. G Zloshchastiev, Quantum entropy of systems described by non-Hermitian Hamiltonians, \href{http://iopscience.iop.org/1742-5468/2016/3/033102}{J. Stat. Mech. 033102 (2016)}.
%
%
%
%
\bibitem{esther1}M. Rigol, V. Dunjko, and M. Olshanii, Thermalization and its mechanism for generic
isolated quantum systems, \href{http://doi.org/10.1038/nature06838}{Nature {\bf 452}, 854 (2008)}.
%
%
\bibitem{esther2}A. Polkovnikov, K. Sengupta, A. Silva, and M. Vengalattore, Colloquium: Nonequilibrium dynamics of closed interacting quantum systems, \href{https://doi.org/10.1103/RevModPhys.83.863}{Rev. Mod. Phys. {\bf 83}, 863 (2011)}.

%
%
\bibitem{nhmeasurementge1}Y. Wu, W. Liu, J. Geng, X. Song, X. Ye, C.-K. Duan, X. Rong, J. Du, Observation of parity-time symmetry breaking in a single-spin system, \href{https://doi.org/10.1126/science.aaw8205}{Science {\bf 364}, 878-880 (2019)}.

%
%
\bibitem{nhmeasurementge2}J. Thingna and P. Talkner, Quantum measurements of sums, \href{https://doi.org/10.1103/PhysRevA.102.012213}{Phys. Rev. A {\bf 102}, 012213 (2020)}.

%
%
\bibitem{nhmeasurementge3}J. Son, P. Talkner, and J. Thingna, Monitoring quantum Otto engines, \href{https://arxiv.org/abs/2105.10665}{arXiv:2105.10665 (2021)}.
%
%
\bibitem{nhSSHer1}K. G. Marris, R. El-Ganainy, and D. N. Christodoulides, Beam dynamics in PT symmetric optical lattices, \href{https://doi.org/10.1103/PhysRevLett.100.103904}{Phys. Rev. Lett. {\bf 100}, 103904 (2008)}.
%
%
%
%
\bibitem{nhSSHer3}Y. Ashida, S. Furukawa, and M. Ueda, Quantum critical behavior influenced by measurement backaction in ultracold gases, \href{https://doi.org/10.1103/PhysRevA.94.053615}{Phys. Rev. A {\bf 94}, 053615 (2016)}.
%
%
\bibitem{nhSSHer4}Y. Ashida, S. Furukawa, and M. Ueda, Parity-time-symmetric quantum critical phenomena, \href{https://doi.org/10.1038/ncomms15791}{Nat. Commun. {\bf 8}, 15791 (2017)}.
%
%
\bibitem{nhSSHer5}G.-Q. Zhang, Y.-P. Wang, and J. Q. You, Dispersive readout of a weakly coupled qubit via the parity-time-symmetric phase transition, \href{https://doi.org/10.1103/PhysRevA.99.052341}{Phys. Rev. A {\bf 99}, 052341 (2019)}.
%
%
\bibitem{metastable}K. Macieszczak, M. Gu\c{t}\v{a}, I. Lesanovsky, and J. P. Garrahan, Towards a theory of metastability in open quantum dynamics, \href{https://doi.org/10.1103/PhysRevLett.116.240404}{Phys. Rev. Lett. {\bf 116}, 240404 (2016)}.
%





\end{references}
\end{document}